\documentclass[aps,pra,reprint,groupedaddress]{revtex4-2}

\usepackage{amsmath}
\usepackage{graphicx}

\begin{document}

\preprint{}

\title{Deterministic Minimum-Leakage Continuous-Variable Quantum Key Distribution with Phase-Conjugated Twin Beams}

\author{Zhenlin Zhao}
\author{Dawei Wang}
\email[]{wangdw9@mail.sysu.edu.cn}
\affiliation{School of Electronics and Information Technology, Sun Yat-sen University, Guangzhou, China}

\begin{abstract}
Minimum-leakage continuous-variable quantum key distribution suppresses Eve's Holevo information by engineering the signal ensemble during state preparation.
The existing symmetric realization relies on Alice-side heralding.
Here we propose an alternative two-mode symmetric realization that is deterministic without heralding.
The minimum-leakage condition of the new protocol requires transmitting two phase-conjugated twin beams (PCTB).
We show that the PCTB protocol is related to the heralding protocol through a common entanglement-based source but corresponds to a different prepare-and-measure decomposition.
The two protocols attain the same secret key rate per transmitted optical mode in the very-large-squeezing limit.
For finite squeezing, however, the PCTB protocol requires approximately 3 dB less squeezing to achieve the same performance.
We further optimize mode-symmetric Gaussian two-mode entangling-cloner attacks and find that correlated ancillary modes provide only a limited advantage over independent attacks under the minimum-leakage condition.
These results establish phase-conjugated twin beams as a deterministic and experimentally appealing route to symmetric minimum-leakage CV-QKD.
\end{abstract}

\maketitle

\section{\label{sec:intro}Introduction}

Quantum key distribution (QKD) enables two distant parties, Alice and Bob, to establish secret keys whose security is guaranteed by the laws of quantum physics \cite{scarani2009security,pirandola2020advances,zhang2024continuous}.
Continuous-variable (CV) QKD encodes information in the field quadratures of optical modes and is attractive because it is compatible with standard telecom components, coherent detection, and Gaussian optical processing \cite{weedbrook2012gaussian,laudenbach2018continuous,zhang2024continuous,usenko2026continuous}.
Recent long-distance, high-rate, and integrated demonstrations further indicate its potential for practical secure optical networks \cite{zhang2020long,hajomer2024continuous,liao2025experimental,wang2025high,wu2026high}.

In the asymptotic collective-attack regime, where Gaussian attacks are optimal for Gaussian-modulated protocols under covariance-matrix constraints \cite{navascues2006optimality,garciapatron2006unconditional}, the reverse-reconciliation secret key rate is bounded by the Devetak-Winter expression $K=\beta I_{AB}-\chi_{BE}$, where $I_{AB}$ is the classical mutual information between Alice and Bob, $\beta$ is the reconciliation efficiency, and $\chi_{BE}$ is the Holevo information bounding Eve's accessible information \cite{grosshans2003reverse,devetak2005distillation}.
This expression highlights two complementary routes toward high-performance CV-QKD: increasing the useful correlations between Alice and Bob, and reducing the information physically leaked to Eve.
The latter is especially appealing because suppressing $\chi_{BE}$ at the state-preparation level can relax the required reconciliation efficiency and reduce the amount of compression needed for privacy amplification \cite{bennett1995generalized,grosshans2003reverse,leverrier2008multidimensional,jacobsen2018complete}.

A foundational example of this idea was introduced by Jacobsen \textit{et al.} \cite{jacobsen2018complete}.
Instead of relying only on classical post-processing to remove Eve's information after transmission, they showed that the Gaussian signal ensemble itself can be engineered such that Eve obtains zero Holevo information in a pure-loss channel.
The protocol uses squeezed states modulated in only one quadrature, with the squeezing variance $V_{\rm sqz}$ and modulation variance $V_{\rm sig}$ satisfying $V_{\rm sqz}+V_{\rm sig}=1$ in shot-noise units (SNU).
Under this minimum-leakage condition, the total noise in the modulated quadrature equals vacuum noise, or 1 SNU.
As a result, the signal remains accessible to Bob, but the information leaked to Eve is eliminated in the loss-only case and minimized in a symmetric noisy channel.
This result establishes minimum leakage as a state-engineering principle for CV quantum communication.

However, the original minimum-leakage protocol is intrinsically one-dimensional: it encodes information in a single quadrature, while the conjugate quadrature remains unmodulated.
This quadrature asymmetry complicates the security analysis.
Winnel \textit{et al.} \cite{winnel2021minimization} showed that, when general asymmetric collective attacks are considered, Eve's information is not necessarily minimized by the same condition that is optimal under symmetric attacks.
If the unmodulated quadrature is bounded only through the Heisenberg uncertainty principle, the resulting security bound can be loose and pessimistic.
To remove this issue, they proposed a symmetric heralding protocol.
In that scheme, Alice prepares two independently modulated squeezed ensembles in orthogonal quadratures, interferes them on a beam splitter, homodynes one output mode, and thereby heralds an effective single-mode squeezed ensemble sent to Bob.
The heralding construction restores symmetry between the two quadratures and asymptotically eliminates information leakage in a pure-loss channel when the corresponding minimum-leakage condition is satisfied.

\begin{figure*}[t]
\centering
\includegraphics[width=\textwidth]{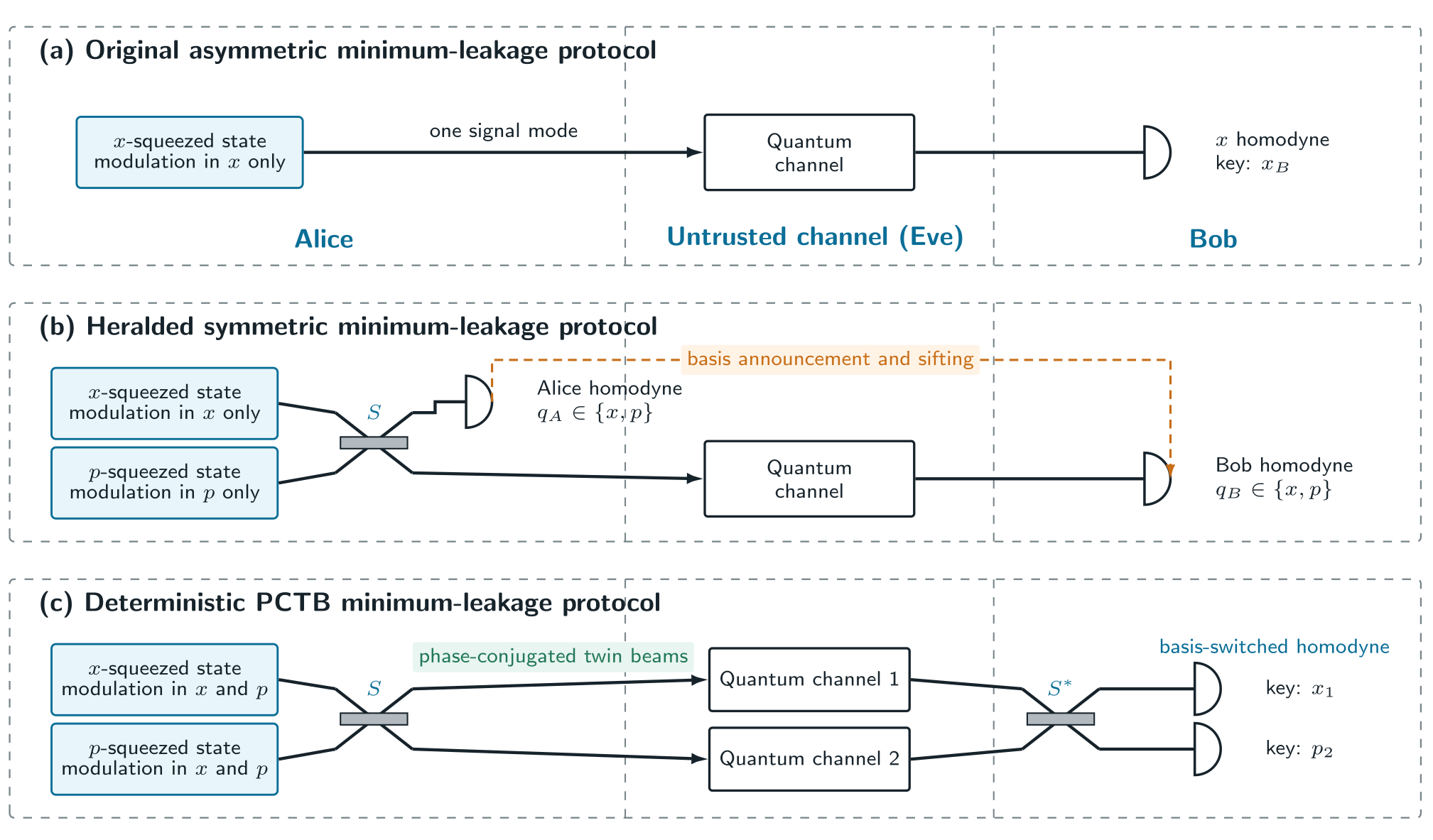}
\caption{Comparison of the original asymmetric \cite{jacobsen2018complete}, heralded symmetric \cite{winnel2021minimization}, and deterministic PCTB minimum-leakage CV-QKD protocols (this work).}
\label{fig:protocol_comparison}
\end{figure*}

The heralding protocol resolves the quadrature-asymmetry problem, but it also raises a natural question: is Alice's heralding measurement fundamentally necessary, or can the two beam-splitter outputs themselves be used directly as a deterministic communication resource?
In this work, we answer this question by proposing a two-mode minimum-leakage CV-QKD protocol without Alice-side heralding.
Alice prepares two independent squeezed-state ensembles, one squeezed in $x$ and the other squeezed in $p$, and combines them on a balanced beam splitter.
Instead of measuring one output and sending the other, Alice transmits both output modes to Bob; the protocol is therefore deterministic and requires no Alice-side heralding.
The two transmitted modes are strongly correlated and form what we call phase-conjugated twin beams (PCTB) because they are compatible with two phase-conjugated Gaussian-modulated coherent beams.

All three protocols are compared in Fig.~\ref{fig:protocol_comparison}.
In the PCTB protocol, the two squeezed-state ensembles before the beam splitter have exactly one SNU of variance in their squeezed quadratures.
This contrasts with the heralding protocol, where Bob's ensemble after Alice's heralding has 1 SNU variance in one quadrature.
A second fundamental difference in state engineering is that the two protocols use different intrinsic squeezed states for modulation.
In particular, the PCTB protocol modulates the squeezed states in \emph{both} quadratures, while the heralding protocol modulates only one quadrature.
We show that the secret key rate (SKR) of the deterministic PCTB protocol is identical to that of the heralding protocol \cite{winnel2021minimization} in the very large squeezing limit.
However, for finite squeezing, the PCTB construction requires about 3 dB less squeezing to achieve the same SKR, making it less demanding on the physical squeezed-state quality.

Multimode structures in state preparation and detection can substantially affect CV-QKD security \cite{usenko2014entanglement,derkach2017continuous}.
The presence of inter-mode correlations also distinguishes the PCTB protocol from previous studies of two-mode attacks in multimode CV-QKD \cite{ottaviani2017gaussian}.
Eve can now adapt the correlations of her ancillary modes to improve her attack, in contrast to the result observed in a multimode coherent-state protocol with an i.i.d. mode setting \cite{ottaviani2017gaussian}.
We show, however, that Eve's quantitative advantage using correlated two-mode attacks remains limited under the minimum-leakage condition.
Consequently, the PCTB construction provides a symmetric two-mode protocol that satisfies the minimum-leakage condition, without switching or heralding on Alice's side.
Moreover, because PCTB states naturally involve paired modes, they are compatible with multiplexed implementations and phase-sensitive processing schemes, which may be useful for high-speed CV-QKD architectures \cite{liao2025experimental,su2023experimental}.

\begin{figure}[t]
\includegraphics[width=\columnwidth]{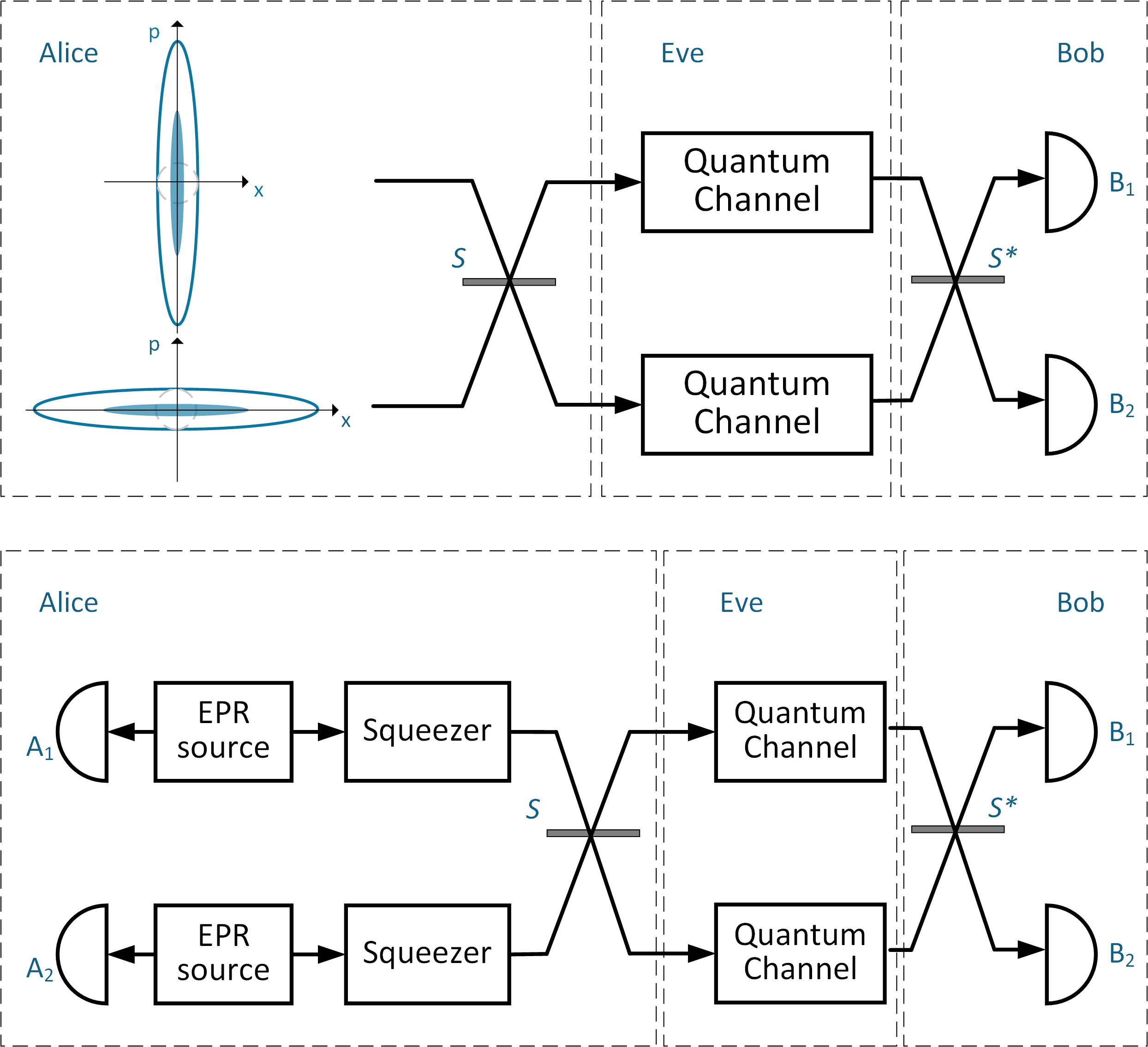}
\caption{\label{fig:pctb_protocol} The prepare-and-measure (upper) and entanglement-based (lower) pictures of the two-mode PCTB CV-QKD protocol.}
\end{figure}

\section{Minimal Leakage Two-Mode Protocol}

The details of the PCTB protocol are illustrated in Fig.~\ref{fig:pctb_protocol}.
Throughout this section, all covariance matrices are normalized to SNU and written in the quadrature ordering $(x,p)$ for one-mode states and $(x_1,p_1,x_2,p_2)$ for two-mode states.

In the prepare-and-measure (PM) picture, Alice prepares two independent Gaussian ensembles of displaced squeezed states with intrinsic covariance matrices $\operatorname{diag}(e^{-2r},e^{2r})$ and $\operatorname{diag}(e^{2r},e^{-2r})$, respectively, where $r>0$ denotes the squeezing parameter.
The classical displacement modulation is applied in both quadratures, with variances chosen such that each averaged ensemble is a squeezed thermal Gaussian state.
The resulting ensemble covariance matrices are
\begin{equation}
\gamma_1=
\begin{pmatrix}
u e^{-2r} & 0 \\
0 & u e^{2r}
\end{pmatrix},
\qquad
\gamma_2=
\begin{pmatrix}
u e^{2r} & 0 \\
0 & u e^{-2r}
\end{pmatrix},
\label{eq:sub_ensemble}
\end{equation}
where $u\geq 1$ is the quadrature variance when $r=0$.
Alice then mixes the two modes on a balanced beam splitter and sends both output modes through the quantum channel.
The covariance matrix of the transmitted two-mode state is
\begin{equation}
\gamma_{A_1A_2}=
\begin{pmatrix}
u\cosh(2r) I_2 & u\sinh(2r) Z \\
u\sinh(2r) Z & u\cosh(2r) I_2
\end{pmatrix},
\label{eq:source_cm}
\end{equation}
where $I_2$ is the $2\times2$ identity matrix and $Z=\operatorname{diag}(1,-1)$.
The state has the same structure as a two-mode squeezed thermal state and is locally phase-insensitive, since each transmitted mode has the same variance in $x$ and $p$, and it is symmetric under exchange of the two modes.

The same protocol can be represented in an entanglement-based (EB) picture.
Alice starts from two independent EPR sources with variance $u$ and applies single-mode squeezing to one mode of each EPR pair, with opposite squeezing directions for the two sources.
She then performs \emph{heterodyne} measurements on the retained modes.
Conditioned on Alice's heterodyne outcomes, the outgoing quantum state after the balanced beam splitter is a displaced two-mode squeezed vacuum (TMSV) state with covariance matrix
\begin{equation}
\gamma_*=
\begin{pmatrix}
\cosh(2r) I_2 & \sinh(2r) Z \\
\sinh(2r) Z & \cosh(2r) I_2
\end{pmatrix}.
\end{equation}
Averaging over Alice's heterodyne outcomes adds the classical Gaussian displacements and recovers the PM ensemble covariance matrix in Eq.~\eqref{eq:source_cm}.

This PM-EB equivalence relies on Alice performing heterodyne measurements, which remotely prepare sub-ensembles modulated in both quadratures.
If Alice instead performs homodyne measurements on the retained EPR modes, the remote preparation yields two sub-ensembles modulated only in the quadratures Alice measures.
This leads to the heralded minimum-leakage construction of Ref.~\cite{winnel2021minimization}, where Alice measures one beam-splitter output and sends the other to Bob.

Note that although the covariance matrix~\eqref{eq:source_cm} is the same as that of modes $A_3$ and $B$ before Alice's heralding measurement in Ref.~\cite{winnel2021minimization}, the two schemes differ in three critical aspects: 1) the intrinsic squeezed state, 2) the minimal-leakage condition, and 3) how the squeezed quadratures are selected.

First, the heralding protocol uses $\operatorname{diag}((ue^{2r})^{-1}, ue^{2r})$ as the intrinsic squeezed state, which results directly from Alice performing homodyne measurements on one EPR mode.
This contrasts with the $\operatorname{diag}(e^{-2r}, e^{2r})$ used in the PCTB protocol, where Alice performs heterodyne measurements on her side of the EPR sources.
Details are further clarified in Appendix~\ref{app:common_source}.

Second, the minimum-leakage condition for the heralding protocol reads \cite{winnel2021minimization}
\begin{equation}
\frac{u}{\cosh{2r}} =
\frac{2u e^{2r}}{e^{4r}+1} = 1 \ ,
\end{equation}
or equivalently $u=\cosh(2r)$, in which case the squeezed quadrature of Bob's mode before the channel, conditioned on Alice's heralding, has exactly one SNU of variance.
Similarly, the minimal-leakage condition for the PCTB protocol requires that the squeezed quadrature of Bob's recovered mode from a loss-only channel has exactly one SNU of variance, which corresponds to the case when the transmitted two-mode ensemble described by the covariance matrix~\eqref{eq:source_cm} lies at the boundary of classical inter-mode correlations \cite{simon2000peres,mandel1995optical}
\begin{equation}
u\cosh(2r)-u\sinh(2r) = u e^{-2r} = 1
\ ,
\label{eq:min_leakage_condition}
\end{equation}
or equivalently $u=e^{2r}$.
Under this condition, the two-mode state is indistinguishable from two coherent beams with phase-conjugated Gaussian modulation, hence the name PCTB.

Finally, in the heralding protocol, the key-generating data are selected by sifting Alice's heralding basis with Bob's homodyne basis.
In the PCTB protocol, Bob first decouples the two received modes into the two squeezed ensembles using a conjugate beam splitter.
Then, he performs basis-switched homodyne detection, i.e., randomly switches between $(x_{B_1},p_{B_2})$ and $(p_{B_1},x_{B_2})$, and uses the data obtained from the squeezed quadrature of each recovered ensemble for key generation.

\begin{figure}[t]
\includegraphics[width=\columnwidth]{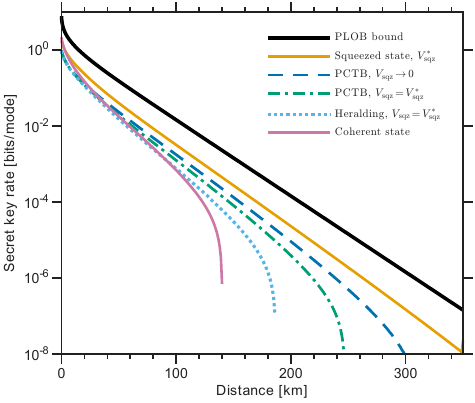}
\caption{\label{fig:skr_comparison} Secret key rate per mode versus transmission distance for the PCTB protocol and standard one-mode benchmarks under independent Gaussian attacks.
The repeaterless bound is the PLOB bound \cite{pirandola2017fundamental}; the coherent-state and squeezed-state baselines follow the standard one-mode protocols \cite{grosshans2002continuous,cerf2001quantum}.
The squeezing variance is optimized at each distance for the one-mode squeezed state protocol, denoted by $V_{\rm sqz}^*$.
The curve of the heralding protocol in the limit $V_{\rm sqz} \to 0$ is not shown because it coincides with that of PCTB in the same condition.
Reconciliation efficiency $\beta=0.95$, excess noise $\xi=0.05$, and ideal detection are assumed.}
\end{figure}

When standard one-mode attacks on the two channels are considered, the SKR per mode versus transmission distance for standard single-mode fiber with a power attenuation coefficient of $0.2$ dB/km is shown in Fig. \ref{fig:skr_comparison}.
The SKR evaluation is detailed in Appendix~\ref{app:skr_evaluation}.
The original one-mode squeezed-state protocol performs better when its squeezing variance is optimized as $V_{\rm sqz}^*$ at each distance.
In the very-large-squeezing limit $V_{\rm sqz}\to 0$, the PCTB and the heralding protocols give the same SKR per mode.

However, for finite squeezing, a performance difference appears when the PCTB and heralding protocols use the same squeezing variance $V_{\rm sqz}$.
In this case, the two protocols use the same underlying state, $\operatorname{diag}(V_{\rm sqz}, V_{\rm sqz}^{-1})$, but generally correspond to different source parameters $(u,r)$ due to their different modulation strategies.
The reason is that although the two protocols have nearly the same $\chi_{BE}$ per mode under the minimum-leakage condition, the PCTB requires approximately 3 dB less squeezing than the heralding protocol to achieve the same $I_{AB}$ per mode (a proof is given in Appendix~\ref{app:finite_squeezing}).

In addition, the PCTB protocol provides a natural two-mode multiplexed key transmission scheme, doubling the total key rate while using the same amount of physical hardware: two modulated squeezed sources and two homodyne detectors.
Note that when $r=0$, the PCTB becomes two identical phase-insensitive Gaussian ensembles, and the protocol reduces to two parallel coherent-state CV-QKD protocols with homodyne detection.

\begin{figure*}[t]
\centering
\includegraphics[width=.9\textwidth]{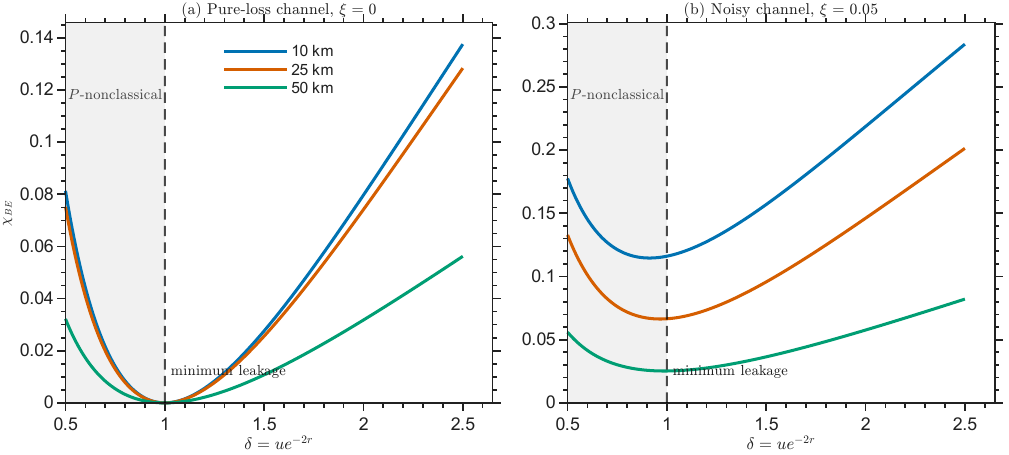}
\caption{\label{fig:minimum_leakage} Holevo information per transmitted optical mode as a function of $\delta=u e^{-2r}$ at a fixed intrinsic squeezing variance $V_{\rm sqz}=e^{-2r}=0.1$ (10 dB).
The vertical line marks the minimum-leakage condition and the Gaussian $P$-classicality boundary $\delta=1$; the shaded region $\delta<1$ is not $P$-classical.
Curves correspond to transmission distances of 10, 25, and 50 km in standard fiber with attenuation $0.2$ dB/km.
(a) Pure-loss channel, $\xi=0$.
(b) Noisy channel, $\xi=0.05$.}
\end{figure*}

To verify the minimum-leakage operating point of the PCTB protocol, we denote the squeezed-quadrature ensemble variance $\delta=u e^{-2r}$.
For a fixed squeezing level, varying $\delta$ changes the Gaussian modulation while leaving the intrinsic squeezed state unchanged.
Figure~\ref{fig:minimum_leakage} shows the Holevo information per transmitted optical mode as a function of $\delta$.
In a pure-loss channel, $\chi_{BE}$ vanishes at $\delta=1$.
In a noisy channel, the same boundary point minimizes the leakage over the $P$-classical region $\delta\geq1$.

Although Bob can decouple the two correlated physical modes using a conjugate beam splitter, keeping them as a two-mode transmitted resource is important.
Conceptually, PCTBs provide a deterministic, fully Gaussian, two-mode version of the minimum-leakage idea: the quadrature-asymmetric one-mode protocol of Ref.~\cite{jacobsen2018complete} is symmetrized without the Alice-side heralding.
Practically, the PCTB source also provides a natural mode-pair structure.
If the two phase-conjugated modes are implemented as frequency-multiplexed modes, they can in principle be processed by phase-sensitive amplification without adding excess noise, or by joint measurement based on phase-sensitive detection \cite{kovalenko2021frequency,su2023experimental,
liao2025experimental,wang2025high}.
This compatibility makes the protocol relevant for high-speed and multiplexed CV-QKD architectures.

Because the signal is transmitted through two correlated physical modes, the appropriate security analysis must go beyond independent single-mode attacks.
In the next section, we therefore study Gaussian two-mode attacks in which Eve injects correlated ancillary modes and optimizes over their inter-mode correlations.

\section{Two-Mode Attacks}

\begin{figure}[t]
\centering
\includegraphics[width=\columnwidth]{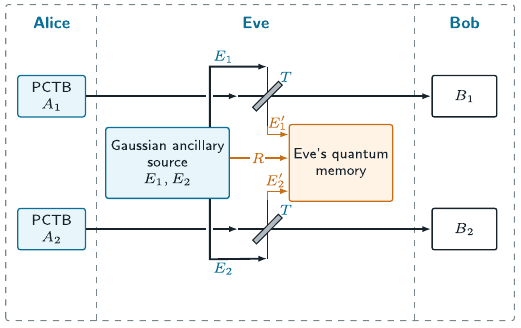}
\caption{\label{fig:entangling_cloner}Two-mode entangling-cloner attack in the transmitted physical-mode basis.
Eve injects the jointly prepared Gaussian ancillary modes $E_1$ and $E_2$ into two beam splitters of identical transmittance $T$, sends the transmitted modes $B_1$ and $B_2$ to Bob, and stores the purification $R$ and the reflected outputs $E'_1$ and $E'_2$ in a quantum memory.}
\end{figure}

We model Eve's correlated attacks with a two-mode entangling cloner, which is illustrated in Fig.~\ref{fig:entangling_cloner}.
Eve replaces the two physical channels with two beam splitters of identical transmittance $T$ and injects two Gaussian ancillary modes through the unused input ports.
Eve is also assumed to retain a purification of the generally mixed ancillary modes.
The two ancillary modes have identical variance $\omega$ with covariance matrix written as
\begin{equation}
\gamma_{E_1E_2}=
\begin{pmatrix}
\omega I_2 & G \\
G & \omega I_2
\end{pmatrix},
\qquad
G=
\begin{pmatrix}
g_1 & 0 \\
0 & g_2
\end{pmatrix},
\label{eq:eve_ancilla_cm}
\end{equation}
where $g_1$ and $g_2$ describe the ancilla's quadrature correlations bounded by the uncertainty condition $\gamma_{E_1E_2}+i\Omega \geq 0$, and the PCTB covariance matrix at the channel output becomes
\begin{equation}
\gamma_{B_1B_2} =
T\gamma_{A_1A_2}+(1-T)\gamma_{E_1E_2}
\ .
\end{equation}
The thermal variance $\omega$ is related to the usual input-referred channel noise by
\begin{equation}
\omega = \frac{T}{1-T}\chi_{\rm line},
\qquad
\chi_{\rm line}=\frac{1-T}{T}+\xi,
\end{equation}
where $\xi$ is the excess noise referred to the channel input.
When $g_1=g_2=0$, the attack reduces to two independent single-mode entangling-cloner attacks.

\begin{figure}
\includegraphics[width=.8\columnwidth]{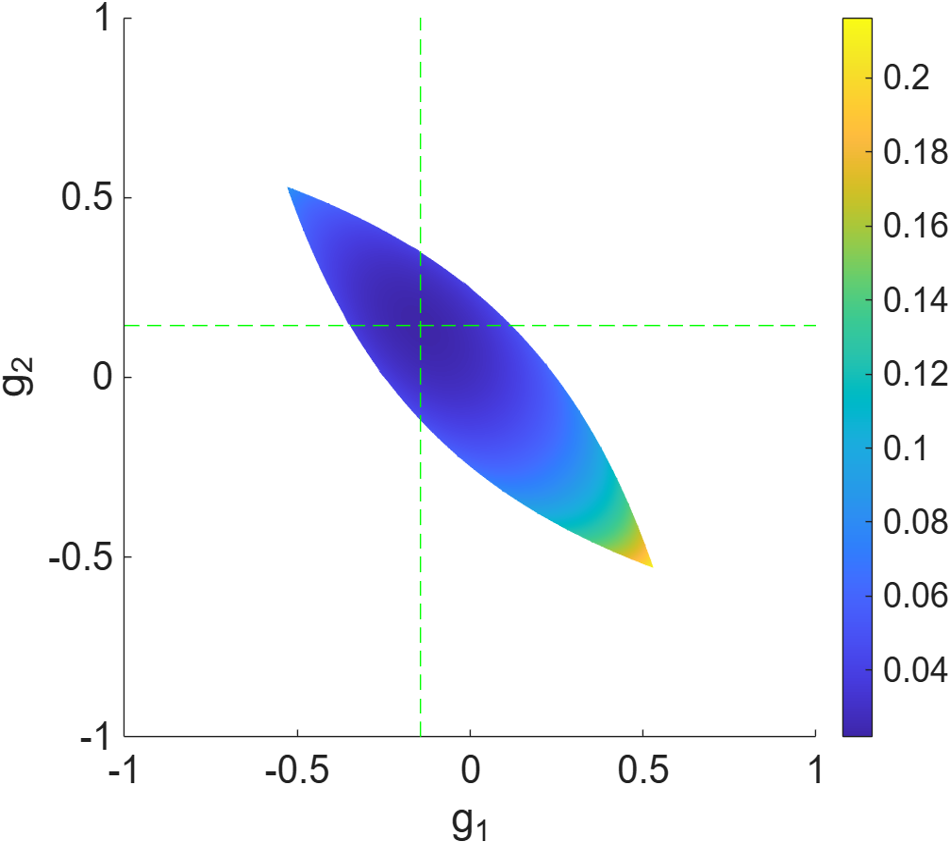}
\caption{\label{fig:attack_correlations} Secret key rate as a function of Eve's ancillary correlations $g_1$ and $g_2$.
The physically allowed region is determined by the uncertainty principle, the color scale denotes the secret key rate, and the dashed lines mark the correlations that minimize the rate.
The minimum key rate is obtained for nonzero correlations in Eve's ancillary modes.
Parameters are $\beta=0.95$, $T=0.4$, $\xi=0.2$, $u=23.5$.}
\end{figure}

\begin{figure*}[t]
\includegraphics[width=.9\textwidth]{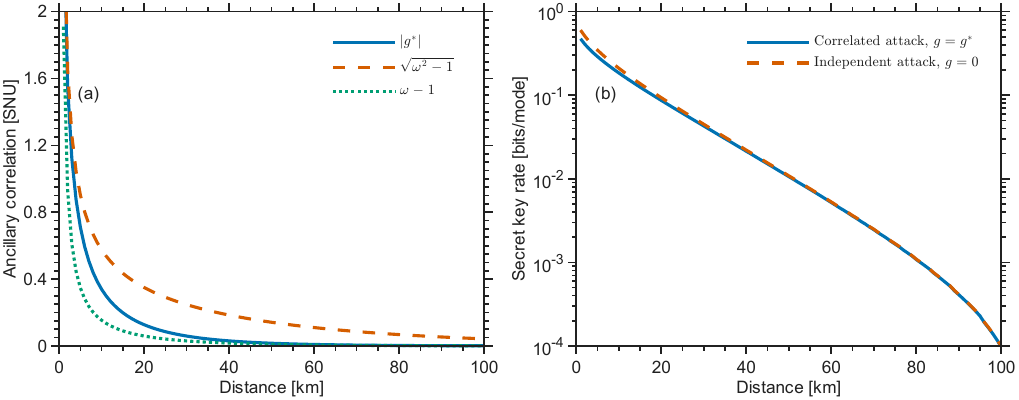}
\caption{\label{fig:optimal_attack} Optimized correlated two-mode attacks with $g_1=-g_2$ as a function of transmission distance.
(a) The optimal ancillary correlation $|g^{*}|$ is compared with the separability boundary $\omega-1$ and the physicality boundary $\sqrt{\omega^2-1}$.
(b) Secret key rates under independent attacks and optimized correlated two-mode attacks.
Parameters are $\beta=0.95$, $\xi=0.09$.}
\end{figure*}

Figure~\ref{fig:attack_correlations} shows the secret key rate obtained by scanning the physically allowed values of $g_1$ and $g_2$.
The optimal points that minimize the SKR are found along the phase-conjugated direction $g_1=-g_2$, which confirms that Eve can gain a small advantage by correlating the two ancillary modes.
The physical effect of the correlated attack becomes transparent after Bob applies the conjugate beam splitter.
For $g_1=-g_2$, Eve's correlated noise is also diagonalized in Bob's decoupled-mode basis.
At the minimum-leakage point, the two decoupled modes acquire effective covariance matrices of the form
\begingroup
\small
\setlength{\arraycolsep}{3.5pt}
\begin{equation}
\gamma'_{1}=
\begin{pmatrix}
T(\chi_{\rm line}+1)-(1-T)g_1 & 0 \\
0 & Tu e^{2r}+T\chi_{\rm line}+(1-T)g_1
\end{pmatrix},
\end{equation}
\begin{equation}
\gamma'_{2}=
\begin{pmatrix}
Tu e^{2r}+T\chi_{\rm line}+(1-T)g_1 & 0 \\
0 & T(\chi_{\rm line}+1)-(1-T)g_1
\end{pmatrix}
\label{eq:effective_decoupled_modes}
\end{equation}
\endgroup
Since $g_1<0$ in the optimal attack, the variance of the originally squeezed quadrature is increased, while that of the anti-squeezed quadrature is reduced.
In this sense, Eve's correlated ancillas \emph{partially unsqueeze} Bob's decoupled modes, subject to the uncertainty-principle constraint on $\gamma_{E_1E_2}$.

Figure~\ref{fig:optimal_attack}(a) further shows how the optimized correlation changes with distance.
For the PCTB operating point considered here, the optimal $|g_1|$ lies between the separability boundary $\omega-1$ and the physicality boundary $\sqrt{\omega^2-1}$.
Thus, Eve's optimal ancillary state is quantum correlated.
In the short-distance regime, the optimum approaches the physicality boundary $|g_1|=\sqrt{\omega^2-1}$, while at long distance it approaches the separability boundary $|g_1|=\omega-1$.
The impact on the key rate is shown in Fig.~\ref{fig:optimal_attack}(b).
The attack improvement is small given the minimum-leakage condition.
The SKR reduction is visible mainly at short distances, while at long distances the two curves become nearly indistinguishable.
This confirms that the PCTB protocol is robust against mode-symmetric Gaussian correlated attacks at long distances.

\begin{figure}[t]
\includegraphics[width=.9\columnwidth]{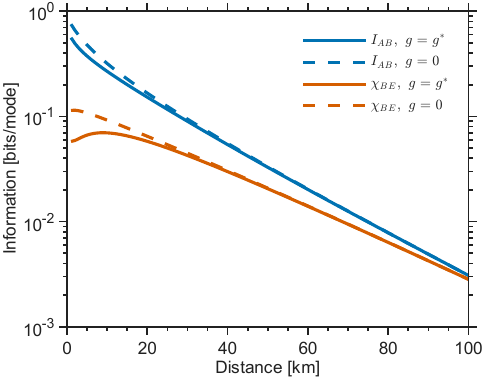}
\caption{\label{fig:attack_information} Alice-Bob mutual information $I_{AB}$ and Eve's Holevo information $\chi_{BE}$ versus transmission distance for the optimized correlated attack $g=g^*$ and the independent attack $g=0$.
The same parameters as in Fig.~\ref{fig:optimal_attack} are used.}
\end{figure}

Figure~\ref{fig:attack_information} separately compares $I_{AB}$ and $\chi_{BE}$ for the optimized correlated attack $g=g^*$ and the independent attack $g=0$, where $g\equiv g_1=-g_2$.
Across the considered distance range, the optimized attack reduces both quantities, but the reduction in $I_{AB}$ dominates.
Thus, the partial unsqueezing lowers the key rate by degrading Alice--Bob correlations rather than by increasing Eve's Holevo information.

\section{\label{sec:con}Conclusion}

We have proposed a deterministic two-mode minimum-leakage CV-QKD protocol based on phase-conjugated twin beams.
The protocol constructs the key by mixing two oppositely squeezed Gaussian ensembles on a balanced beam splitter and sending both output modes to Bob.
At the minimum-leakage point, the squeezed-quadrature ensemble variance is one shot-noise unit, which eliminates Eve's Holevo information in a pure-loss channel and strongly suppresses it in a noisy Gaussian channel.
We also clarify that both PCTB and heralding protocols can be viewed as different prepare-and-measure decompositions of a common entanglement-based Gaussian source, depending on the type of measurement Alice should perform in the EB-equivalent scheme.
The two schemes attain the same secret key rate per transmitted optical mode in the very-large-squeezing limit.
For finite squeezing, however, the PCTB protocol requires less squeezing to obtain the same secret key rate.
The PCTB construction allows Eve to slightly improve her attack by injecting correlated ancillary modes; the resulting key-rate reduction is small near the minimum-leakage point and becomes negligible at long distances.
The PCTB construction therefore provides a deterministic two-mode implementation of the minimum-leakage principle and is naturally compatible with mode-pair architectures such as frequency multiplexing and phase-sensitive optical processing.

\begin{acknowledgments}
This work was supported in part by the National Natural Science Foundation of China under Grant 62475297 and in part by the Guangdong Basic and Applied Basic Research Foundation under Grant 2024B1515020004.
\end{acknowledgments}

\appendix

\section{Common entanglement-based source and prepare-and-measure decompositions}
\label{app:common_source}

In this Appendix, we clarify the relation between the deterministic PCTB protocol and the heralded protocol.
We use the quadrature ordering $(x_1,p_1,x_2,p_2)$ and define the balanced beam splitter as
\begin{equation}
S_{\rm BS}=\frac{1}{\sqrt{2}}
\begin{pmatrix}
I_2 & I_2\\
-I_2 & I_2
\end{pmatrix}.
\end{equation}

Consider two independent EPR sources with variance $u$.
Alice applies single-mode squeezing to one mode of each EPR pair, with opposite squeezing directions for the two sources.
The two squeezed modes are then mixed by $S_{\rm BS}$.
The averaged covariance matrix of the two outgoing modes is
\begin{equation}
\gamma_{\rm out}
=
\begin{pmatrix}
u\cosh(2r)I_2 & u\sinh(2r)Z\\
u\sinh(2r)Z & u\cosh(2r)I_2
\end{pmatrix},
\label{eq:app_common_out}
\end{equation}
where $Z=\operatorname{diag}(1,-1)$.
This is the common two-mode Gaussian source shared by the present PCTB protocol and the heralded construction before Alice's heralding measurement.

The difference between the two protocols is the prepare-and-measure decomposition of this same ensemble.
If Alice performs heterodyne measurements on the retained EPR modes, the conditional state before classical averaging is
\begin{align}
\gamma_{\rm core}^{\rm het}
&=
S_{\rm BS}
\left[
\begin{pmatrix}
e^{-2r} & 0\\
0 & e^{2r}
\end{pmatrix}
\oplus
\begin{pmatrix}
e^{2r} & 0\\
0 & e^{-2r}
\end{pmatrix}
\right]
S_{\rm BS}^{T}\\
&=
\begin{pmatrix}
\cosh(2r)I_2 & \sinh(2r)Z\\
\sinh(2r)Z & \cosh(2r)I_2
\end{pmatrix}.
\end{align}
Averaging over the heterodyne outcomes adds the Gaussian displacement modulation
\begin{equation}
(u-1)
\left[
\begin{pmatrix}
e^{-2r} & 0\\
0 & e^{2r}
\end{pmatrix}
\oplus
\begin{pmatrix}
e^{2r} & 0\\
0 & e^{-2r}
\end{pmatrix}
\right]
\end{equation}
before the beam splitter and therefore recovers Eq.~\eqref{eq:app_common_out}.
This is the PM decomposition used by the PCTB protocol.

If Alice instead performs homodyne measurements on the retained EPR modes, the conditional state before the beam splitter is
\begin{align}
\gamma_{1,{\rm core}}^{\rm hom}
=
\begin{pmatrix}
(u e^{2r})^{-1} & 0\\
0 & u e^{2r}
\end{pmatrix},
\\
\gamma_{2,{\rm core}}^{\rm hom}
=
\begin{pmatrix}
u e^{2r} & 0\\
0 & (u e^{2r})^{-1}
\end{pmatrix}.
\end{align}
The missing variance is supplied only in the squeezed quadrature of each sub-ensemble.
Thus the modulation variances are
\begin{align}
V_{{\rm mod},1}^{x}
=
u e^{-2r}-(u e^{2r})^{-1},
\\
V_{{\rm mod},2}^{p}
=
u e^{-2r}-(u e^{2r})^{-1},
\end{align}
while the conjugate quadratures are not modulated.
After the balanced beam splitter, this homodyne-based decomposition gives the same averaged covariance matrix in Eq.~\eqref{eq:app_common_out}.
This decomposition underlies the heralded minimum-leakage protocol in Ref.~\cite{winnel2021minimization}.

\section{Secret-key-rate evaluation from covariance matrices}
\label{app:skr_evaluation}

In this Appendix, we summarize the covariance-matrix calculation of the mutual and Holevo information used throughout the numerical analysis.
The formulas assume ideal detection, as in the main text.

After the two physical channels and Bob's conjugate beam splitter, the covariance matrix of Alice's two retained modes and Bob's two recovered modes is written as
\begin{equation}
\Gamma_{AB}
=
\begin{pmatrix}
A&C\\
C^{T}&B
\end{pmatrix},
\label{eq:app_gamma_ab}
\end{equation}
where $A$, $B$, and $C$ are $4\times4$ matrices in the mode ordering $(A_1,A_2,B_1,B_2)$.
Alice's heterodyne measurement adds one SNU to the measured quadrature variance.
For a key-generating quadrature $q\in\{x,p\}$, the corresponding classical covariance matrix can therefore be written as
\begin{equation}
\Sigma_q
=
\begin{pmatrix}
a_q+1&c_q\\
c_q&b_q
\end{pmatrix},
\end{equation}
where $a_q$, $b_q$, and $c_q$ are obtained from Eq.~\eqref{eq:app_gamma_ab}.
The Gaussian mutual information is
\begin{align}
I(A_q:B_q)
&=
\frac{1}{2}\log_2
v_q \equiv
\frac{1}{2}\log_2
\frac{a_q+1}{a_q+1-c_q^2/b_q}
\ .
\label{eq:app_mutual_information}
\end{align}
For the joint key setting $(x_{B_1},p_{B_2})$, the mutual information of one PCTB block is
\begin{equation}
I_{AB}^{(2)}
=
\frac{1}{2}\log_2 v_{x_1}
+
\frac{1}{2}\log_2 v_{p_2}.
\label{eq:app_iab_pctb}
\end{equation}

For reverse reconciliation, Eve's Holevo information about Bob's joint key variable $b=(x_{B_1},p_{B_2})$ is
\begin{equation}
\chi_{BE}^{(2)}
=
S(E)-S(E|b).
\end{equation}
Because Eve purifies the state shared by Alice and Bob, the two entropies can be evaluated from $\Gamma_{AB}$ and Alice's covariance matrix conditioned on Bob's homodyne outcomes.
Defining
\begin{equation}
\Pi=X\oplus P,
\quad
X=\operatorname{diag}(1,0),
\quad
P=\operatorname{diag}(0,1),
\end{equation}
the conditional covariance matrix is
\begin{equation}
\Gamma_{A|b}
=
A-C\Pi(\Pi B \Pi)^{\rm MP}\Pi C^{T},
\label{eq:app_conditional_cm}
\end{equation}
where ${\rm MP}$ denotes the Moore--Penrose pseudoinverse.
Let $\{\nu_k\}_{k=1}^{4}$ and $\{\widetilde{\nu}_j\}_{j=1}^{2}$ be the symplectic eigenvalues of $\Gamma_{AB}$ and $\Gamma_{A|b}$, respectively.
The Holevo information of one two-mode block is then
\begin{equation}
\chi_{BE}^{(2)}
=
\sum_{k=1}^{4}h(\nu_k)
-
\sum_{j=1}^{2}h(\widetilde{\nu}_j),
\label{eq:app_holevo_pctb}
\end{equation}
where
\begin{equation}
h(\nu)
=
\frac{\nu+1}{2}\log_2\frac{\nu+1}{2}
-
\frac{\nu-1}{2}\log_2\frac{\nu-1}{2}.
\end{equation}

The PCTB rate plotted in this work is normalized per transmitted optical mode.
Bob uses the joint key setting with probability $p_{\rm key}=1/2$, while the complementary setting $(p_{B_1},x_{B_2})$ is used for parameter estimation.
The plotted rate is therefore
\begin{equation}
K_{\rm PCTB}^{\rm per\ mode}
=
\frac{p_{\rm key}}{2}
\max\left[\beta I_{AB}^{(2)}-\chi_{BE}^{(2)},0\right]
.
\label{eq:app_pctb_rate_normalization}
\end{equation}

For the heralding protocol, Alice's heralding measurement leaves a three-mode covariance matrix comprising her two retained modes and Bob's transmitted mode.
The Holevo information is evaluated from its three symplectic eigenvalues and the two symplectic eigenvalues of Alice's state conditioned on Bob's homodyne outcome.
For either retained heralding branch,
\begin{align}
I_{AB}^{\rm H}
&=
\frac{1}{2}\log_2 v_{\rm H},
\\
\chi_{BE}^{\rm H}
&=
\sum_{k=1}^{3}h(\nu_k^{\rm H})
-
\sum_{j=1}^{2}h(\widetilde{\nu}_j^{\rm H}).
\end{align}
After the heralding and homodyne bases are sifted, its rate per transmitted optical mode is
\begin{equation}
K_{\rm H}
=
\frac{1}{2}
\max\left[\beta I_{AB}^{\rm H}-\chi_{BE}^{\rm H},0\right].
\label{eq:app_heralding_rate_normalization}
\end{equation}

\section{Finite-squeezing resource comparison}
\label{app:finite_squeezing}

In this Appendix, we compare the physical squeezing resources required by the PCTB and heralded protocols.
We quantify the squeezing resource by its anti-squeezed variance
\begin{equation}
Q=\frac{1}{V_{\rm sqz}},
\end{equation}
where $V_{\rm sqz}$ is the squeezed-quadrature variance of the intrinsic single-mode squeezed state.

The variance ratio for the PCTB protocol is
\begin{equation}
v_{\rm P} = \frac{V_B}{V_{B|A}} = Q_{\rm P}
\label{eq:app_v_pctb}
\end{equation}
such that the Shannon information is computed as
\begin{equation}
I_{AB} = \frac{1}{2} \log_2(v_{\rm P}) \ .
\end{equation}

The variance ratio for the heralded protocol is
\begin{equation}
v_{\rm H} = \frac{V_B}{V_{B|A}} =
\frac{1}{2}
\left(
Q_{\rm H}+\frac{1}{Q_{\rm H}}
\right).
\label{eq:app_v_her}
\end{equation}

Equations~\eqref{eq:app_v_pctb} and \eqref{eq:app_v_her} show that the two protocols are not equivalent at fixed physical squeezing.
For the same anti-squeezed variance $Q$, the PCTB protocol gives $v_{\rm P}=Q$, whereas the heralded protocol gives $v_{\rm H}=(Q+Q^{-1})/2$.
The two protocols have the same $I_{AB}$ only when
\begin{equation}
Q_{\rm P}
=
\frac{1}{2}
\left(
Q_{\rm H}+\frac{1}{Q_{\rm H}}
\right).
\end{equation}
Solving for $Q_{\rm H}$ gives
\begin{equation}
Q_{\rm H}
=
Q_{\rm P}
+
\sqrt{Q_{\rm P}^{2}-1}.
\label{eq:app_q_relation}
\end{equation}
Therefore, the heralded protocol requires more physical squeezing to achieve the same ideal Alice--Bob correlation and the same secret key rate under the minimum-leakage condition.
The squeezing penalty is
\begin{equation}
\Delta_{\rm dB}
=
10\log_{10}
\left(
\frac{Q_{\rm H}}{Q_{\rm P}}
\right)
=
10\log_{10}
\left(
1+\sqrt{1-\frac{1}{Q_{\rm P}^{2}}}
\right),
\end{equation}
which approaches $10\log_{10}2 \simeq 3~{\rm dB}$ in the large-squeezing limit, $Q_{\rm P} \to \infty$.

\bibliography{reference}

\end{document}